\documentclass[lettersize,journal]{IEEEtran}
\usepackage{amsmath,amsfonts}
\usepackage{algorithmic}
\usepackage{algorithm}
\usepackage{array}
\usepackage[caption=false,font=normalsize,labelfont=sf,textfont=sf]{subfig}
\usepackage{textcomp}
\usepackage{stfloats}
\usepackage{url}
\usepackage{verbatim}
\usepackage{graphicx}
\usepackage{cite}
\usepackage{makecell}
\usepackage{booktabs}
\usepackage{multirow}
\begin{document}

\title{TopoCryptoEEG: End-to-End Encrypted Framework for Brainprint Authentication}

\author{Qingyu Zeng, Lei Sun, Shuai Zhang, Xiuqing Mao, Ting Zhang, Min Zhao, and Wenle Dong%
\thanks{\textit{(Corresponding authors: Lei Sun and Shuai Zhang.)}}%
\thanks{The authors are with the Information Engineering University, Zhengzhou 450001, China (e-mail: sl20210221@163.com; xdzhangs@163.com).}%
}



\maketitle

\begin{abstract}
Brainprint authentication enables continuous identity verification, but electroencephalography (EEG) data remain privacy-sensitive throughout the authentication pipeline. Existing feature-level encrypted authentication (FLEA) schemes encrypt extracted features while relying on a trusted client. We show that a compromised client with access to plaintext query features and continuous authentication scores can exactly recover a registered template after a finite number of interactions. We propose TopoCryptoEEG, to the best of our knowledge, the first homomorphic-encryption-based framework for end-to-end encrypted brainprint authentication. The client encrypts preprocessed EEG data, while the server performs all subsequent operations, including feature extraction, in the ciphertext domain, removing the known-query-feature condition required by the direct linear template-recovery attack analyzed for FLEA. However, the spatiotemporal structure of EEG and the computational cost of homomorphic encryption challenge real-time authentication. We introduce Multi-Window Topology-Aware Encoding (MWTE) and a Topographic Branch Model with Linear Fusion (TopoFuse-MLP), which adapt EEG representations to encrypted computation and limit multiplicative depth. We further propose Query–Patch Coordinate Packing (QPCP) and Lane-Preserving BSGS Evaluation (LP-BSGS) to reduce ciphertext operations. Experiments on standard EEG datasets show that TopoCryptoEEG maintains high authentication accuracy while achieving CPU homomorphic-computation latencies of 1–2 s and GPU latencies near 14 ms. Compared with the strongest baseline considered (EER, IMWUT 2025), TopoCryptoEEG achieves an average speedup exceeding 18×, demonstrating practical feasibility.
\end{abstract}


\section{Introduction}
\IEEEPARstart{B}{rainprint} authentication, also referred to as EEG-based authentication, is a promising biometric technology for applications such as continuous identity verification, device unlocking, and access control. EEG, however, is highly sensitive neural data. Its disclosure may lead to privacy violations and identity theft, and may reveal a user’s health or emotional state. Unlike a password, EEG is a non-renewable raw biometric: once exposed, the underlying physiological patterns cannot be reset \cite{ref1,ref2}. The consequences of an EEG breach are therefore particularly serious, making stronger protection necessary. Such protection is also required by laws governing the collection, storage, and use of neural data, including Colorado’s Protect Privacy of Biological Data Act \cite{ref3}.

A brainprint-authentication system typically consists of a data-acquisition client, a feature extractor, and a server-side reference-feature database. The feature extractor maps EEG data to a compact identity-feature vector, which is encrypted and stored on the server, as illustrated in Fig. 1. The system faces threats from two directions: a malicious client may attempt to infer neural data from the server’s reference database, while a malicious server may seek sensitive information from client-provided data \cite{ref4,ref5}. To mitigate these threats, prior work, such as Zhang et al. \cite{ref6}, has used homomorphic encryption (HE) to protect authentication features, as shown in the upper part of Fig. 1. To reduce encrypted-computation overhead, these schemes assume that the local client is trusted: feature extraction is performed in plaintext on the client, after which the features are encrypted and matched on the server in the ciphertext domain. We refer to this design as feature-level encrypted authentication (FLEA). The client consequently holds both the feature extractor and plaintext features. If the client is compromised, an attacker can launch a template-recovery attack \cite{ref7,ref8}. Protecting only the feature-matching stage is therefore insufficient; an end-to-end encrypted framework that covers the full pipeline is needed.
\begin{figure}[!t]
\centering
\includegraphics[width=3.0in]{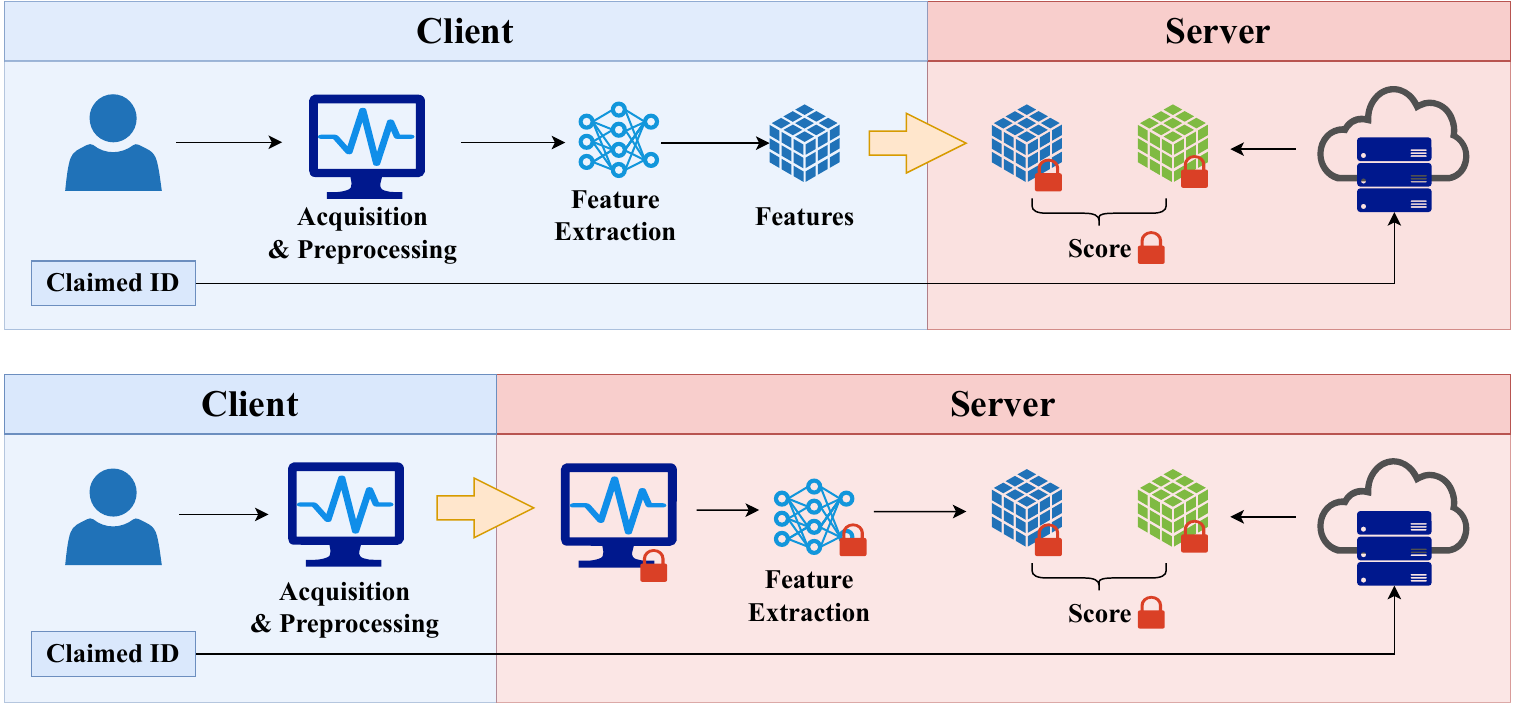}
\caption{Secure brainprint-authentication systems.Top: Feature-level encryption protects the extracted features but relies on a trusted client. Bottom: TopoCryptoEEG encrypts the preprocessed EEG input and performs feature extraction and matching in the ciphertext domain.}
\label{fig_1}
\end{figure}
Research on image classification, emotion recognition, speech analysis, and face recognition has demonstrated the feasibility of end-to-end encrypted inference \cite{ref9,ref10,ref11,ref12}. These approaches use HE to make data “usable but not visible”: the client encrypts the input, and the server performs the main inference operations without accessing plaintext. This paradigm motivates our design, in which the client encrypts the acquired EEG while the server performs the main authentication operations without seeing plaintext. However, encrypted schemes designed for two-dimensional images or one-dimensional sequences cannot be directly applied to EEG, which has multichannel spatial topology and multi-window temporal structure. A dedicated end-to-end encrypted-computation framework is thus needed for EEG data and the brainprint-authentication pipeline.

Prior studies have separately established the feasibility of encrypted EEG analysis, electrode-topology modeling, and encrypted inference. Liu et al., Popescu et al., and Abou Harb et al. applied HE to EEG classification or anomaly detection \cite{ref13,ref14,ref15}, while Dong et al. incorporated electrode topology into privacy-preserving depression diagnosis \cite{ref16}. These studies target classification or diagnosis and do not cover the complete brainprint-authentication pipeline. More recently, Cha et al. interleaved two feature views from personalized collaborative training in a single ciphertext to enable parallel computation in encrypted emotion recognition \cite{ref9}, demonstrating the value of designing ciphertext layouts around data organization. Their method, however, early-fuses multimodal features into a unified vector and organizes the ciphertext for encrypted collaborative training and emotion classification; it does not address electrode-topology-to-slot mapping or brainprint-template distance computation.

These observations point to the need for an encrypted-computation method that is co-designed with the authentication pipeline. General-purpose SIMD packing, tensor layouts, and BSGS-based linear transformations can reduce the cost of individual operators \cite{ref17,ref18,ref19,ref20,ref21,ref22}. Under a complex authentication workload, however, incompatible slot organizations across stages can disrupt data correspondence and necessitate data rearrangement and repeated evaluation, offsetting the gains from operator-level optimization. The feature-extraction network must also preserve discriminative power while strictly limiting ciphertext multiplicative depth. Optimizing a single linear operator or the model alone is therefore insufficient to resolve the efficiency bottlenecks of the complete authentication path.

To address these challenges, we propose TopoCryptoEEG, an end-to-end encrypted brainprint-authentication framework that tightly co-designs input representations and ciphertext evaluation. As shown in the lower part of Fig. 1, the client encrypts publicly preprocessed EEG data and sends it to the server. The server derives encrypted identity features, matches them against encrypted registration templates, and returns an encrypted score. The client decrypts the score and accepts or rejects the claim according to its sign. This design moves the main cloud-side authentication operations into the ciphertext domain, protecting online inputs and intermediate identity features. The main contributions are as follows:
\begin{list}{}{}
\item{\textbf{1. End-to-end encrypted brainprint authentication.} We show that a registered template can be uniquely recovered from exact scores and known query embeddings when their difference matrix has full rank. Experiments demonstrate low-error recovery under the evaluated FLEA settings. We therefore propose TopoCryptoEEG, which, to the best of our knowledge, is the first end-to-end encrypted brainprint-authentication framework to perform feature extraction, template matching, score computation, and threshold centering in the ciphertext domain. Because the client no longer obtains plaintext features, the plaintext query embeddings required for the known-coefficient linear attack are unavailable to the client.} 
\item{\textbf{2. Efficient encrypted computation tailored to EEG data.} We introduce MWTE to align the EEG representation with homomorphic-computation layouts, develop TopoFuse-MLP to limit encrypted-inference depth, and use QPCP and LP-BSGS to reduce computation. These designs retain cross-region fusion while reducing multiplicative depth and ciphertext-computation latency.}
\item{\textbf{3. Feasibility and effectiveness evaluation.} We evaluate the framework on two standard EEG datasets, HGD and BCICIV-2a. The results show that TopoCryptoEEG maintains high accuracy while achieving CPU ciphertext-computation latencies of 1–2 s and GPU latencies close to 14 ms, demonstrating the feasibility of near-real-time secure authentication.}
\end{list}
The remainder of this paper is organized as follows. Section II reviews brainprint authentication, privacy protection, and homomorphic-encryption-based neural-network inference. Section III presents the framework and its security analysis. Section IV details the proposed methods. Section V describes the experimental design and results. Section VI concludes the paper.

\section{Related Work}
\subsection{Brainprint Authentication and Privacy Protection}

Neuroscience studies have shown that EEG exhibits individual-specific and relatively stable neurophysiological patterns, making it a suitable biometric for identity authentication \cite{ref23,ref24,ref25}. Marcel and Millán used EEG with maximum a posteriori model adaptation for personal authentication, demonstrating the feasibility of distinguishing individuals with this modality \cite{ref26}. La Rocca et al. further investigated individual distinctiveness in EEG spectral-coherence connectivity, showing that cross-channel relationships can also carry identity-related information \cite{ref27}. These EEG-based biometric patterns are commonly termed brainprints \cite{ref24}. Unlike traditional biometrics such as fingerprints and faces, however, EEG has distinctive privacy implications: it may encode health, cognitive ability, emotional responses, neurological conditions, and task states. Existing EEG privacy-protection research has focused primarily on stored templates or distributed training. For example, Bidgoly et al. \cite{ref28}, Wang et al. \cite{ref29,ref30}, and the PolyCosGraph system \cite{ref31} use irreversible transformations to generate revocable templates and mitigate storage leakage. Federated learning has also been explored for EEG diagnosis to reduce centralized storage of raw signals \cite{ref32}. These approaches protect stored templates, distributed training, or sensitive attributes in learned representations, but their objectives differ from the requirement in online authentication to prevent the cloud from accessing plaintext EEG representations. HE offers a means of enabling cloud-side computation while keeping data “usable but not visible.”

Zhang et al. \cite{ref6} proposed a secure brainprint-recognition system to protect brainprint data. To reduce the high computational cost of HE, their scheme keeps the feature extractor on the client and in plaintext, encrypting only the extracted features for server-side matching. This design relies on the client-side trust boundary and may expose templates to score-based recovery when query embeddings are known, as analyzed in Section III-A.

If feature extraction and authentication matching are both to be performed by the server, an end-to-end encrypted framework is required. Precedents exist in other modalities. Boddeti \cite{ref33} studied secure face-feature matching, and Ao and Boddeti later proposed CryptoFace, which brings both face-feature extraction and matching into the encrypted pipeline \cite{ref11}. CryptoNets performs neural-network classification on encrypted images and returns encrypted predictions \cite{ref10}. Zhang et al. used a deep polynomial network for encrypted speech recognition, with subsequent decoding performed collaboratively by the cloud and the client \cite{ref12}. These studies show that computation protection can be extended from template matching to feature extraction, but EEG data structure and the brainprint-authentication workflow require a dedicated encrypted-computation method.

\subsection{Homomorphic Encryption-Based Neural Network Inference}

Homomorphic encryption (HE) allows a server to perform addition and multiplication on encrypted inputs without decrypting them, offering a way to protect EEG processing. CKKS supports approximate real-number arithmetic and packed encoding, making it suitable for linear transformations and low-degree polynomial activations in neural networks \cite{ref34}. Prior work has applied HE to EEG analysis: Liu et al. studied encrypted EEG classification \cite{ref13}; Popescu et al. investigated encrypted seizure detection and alcohol-dependence prediction \cite{ref14}; Dong et al. combined electrode topology with encrypted depression diagnosis \cite{ref16}; and Abou Harb et al. implemented encrypted dyslexia classification \cite{ref15}. These studies establish the feasibility of encrypted EEG inference, but their computational pipelines are designed primarily for a single classification output. Brainprint authentication has a fundamentally different computational objective: it requires identity representations, template matching, and aggregation of scores across multiple windows.

Cha et al. proposed EER, an encrypted emotion-recognition framework \cite{ref9}. In personalized collaborative training, EER places two feature views in even and odd slots, respectively, and doubles the within-view rotation stride so that both views can be processed in parallel in one ciphertext. This design illustrates the role of slot organization in data-aware encrypted computation. Although the Hide-and-Seek dataset used in EER includes EEG, the model early-fuses multimodal inputs into a feature vector, flattening electrode-space topology. Moreover, EER targets emotion or stress classification and does not address brainprint-template distances or cross-window score reduction.

General-purpose HE inference methods optimize ciphertext layouts, linear transformations, and polynomial activations to reduce evaluation cost and multiplicative depth \cite{ref10,ref17,ref18,ref19,ref20,ref21,ref22}. EEG decoding models may also combine convolutional and Transformer modules, as illustrated by EEG Conformer \cite{ref35}. However, existing methods do not jointly preserve query-window and electrode-patch correspondence through encrypted feature extraction, template matching, and cross-window aggregation. We therefore co-design the representation, model, and ciphertext layout for brainprint authentication.

\section{Security Analysis}
This section first presents the threat model and security analysis, and finally describes the overall system architecture.

\subsection{Threat Model and Security Analysis}

We consider a two-party, semi-honest model consisting of a client $\mathcal{C}$ and a cloud server $\mathcal{S}$. Both parties follow the protocol but may analyze their own data and interaction transcripts; at most one party is corrupted, and the parties do not collude. Although both parties execute the prescribed protocol, either may attempt to infer additional information by analyzing data received from the other party \cite{ref36}.

A potentially adversarial server may try to infer sensitive client EEG from the ciphertexts it observes. We use the CKKS homomorphic-encryption scheme, whose security is based on the Ring Learning with Errors (RLWE) assumption, and rely on its semantic security (IND-CPA) under appropriate parameter settings \cite{ref34}. Assuming that the client’s secret key is not compromised and side-channel attacks are excluded, the server cannot recover the client input or its encrypted features in probabilistic polynomial time from ciphertexts alone. Thus, input confidentiality against the server reduces to the established cryptographic security of CKKS.

A potentially adversarial client may issue multiple valid requests for a target identity and receive the corresponding decrypted continuous scores, but cannot inspect server state, forge protocol messages, or access another key domain. We compare two architectures. In FLEA, the client knows the plaintext query embeddings and the server returns a continuous distance score. In TopoCryptoEEG, the client performs only public preprocessing and input encryption; the feature-extractor parameters and query embeddings remain server-side and encrypted. We therefore analyze whether the known-coefficient template-recovery attack against FLEA remains applicable.

\textbf{Theorem 1 (Template Recovery under FLEA)}

Assume that the client can obtain exact continuous scores for a fixed registered template and knows all query embeddings for each request. If there are $d+1$ protocol-compliant queries whose mean-query-embedding difference vectors form a full-rank matrix, then the client can uniquely recover the target template $e^T$ using at most $d+1$ queries.

\textbf{Proof:}

Define the mean query embedding for request $i$ as follows:
\begin{equation}
    \label{eq:mean-query}
    \bar{e}_i^Q = \frac{1}{q} \sum_{a=0}^{q-1} e_{i,a}^Q.
\end{equation}

Expanding the squared Euclidean distance:
\begin{equation}
    \label{}
    S_{u,i} = \sum_{a=0}^{q-1} \|e_{i,a}^Q\|_2^2 - 2q(\bar{e}_i^Q)^T e^T + q\|e^T\|_2^2 - q\tau.
\end{equation}

Let:
\begin{equation}
    \label{}
    a_i = \sum_{a=0}^{q-1} \|e_{i,a}^Q\|_2^2, \quad c = q\|e^T\|_2^2 - q\tau.
\end{equation}

Then:
\begin{equation}
    \label{}
    S_{u,i} - a_i = -2q(\bar{e}_i^Q)^T e^T + c. 
\end{equation}

Because $c$ is constant across all queries for the same target identity, take request 0 as a reference and subtract its equation from that for request $i$:
\begin{equation}
    \label{eq:delta-e^t}
    \begin{aligned} \Delta_i &= (S_{u,i} - a_i) - (S_{u,0} - a_0) \\ &= -2q(\bar{e}_i^Q - \bar{e}_0^Q)^T e^T. \end{aligned}
\end{equation}

Define:
\begin{equation}
    \label{}
    v_i = \bar{e}_i^Q - \bar{e}_0^Q.
\end{equation}

For the remaining $d$ queries, construct:
\begin{equation}
    \label{eq:matrix v&delta}
    V = \begin{bmatrix} (v_1)^T \\ (v_2)^T \\ \vdots \\ (v_d)^T \end{bmatrix} \in \mathbb{R}^{d \times d}, \quad \Delta = \begin{bmatrix} \Delta_1 \\ \Delta_2 \\ \vdots \\ \Delta_d \end{bmatrix}.
\end{equation}

It follows from \eqref{eq:delta-e^t} that:
\begin{equation}
    \label{}
    \Delta = -2q V e^T.
\end{equation}

By the theorem's assumption, $V$ is full rank and therefore has a unique solution:
\begin{equation}
    \label{}
    e^T = -\frac{1}{2q} V^{-1} \Delta.
\end{equation}

Thus, under the full-rank condition, $d+1$ score queries suffice: one reference query and $d$ linearly independent difference queries. For $d=32$, this requires 33 queries in the ideal exact-score setting. A natural objection is that homomorphic encryption introduces numerical error; could this error disrupt the solution of the linear system? More specifically, suppose the decrypted scores have bounded numerical error $|\xi_{i}| \leq \eta$, and let $\hat{e}^{T}$ be the template recovered from the perturbed scores. Then:
\begin{equation}
    \label{eq:sigma}
    \|\hat{e}^T - e^T\|_2 \leq \frac{\eta \sqrt{d}}{q \sigma_{\min}(V)}.
\end{equation}

Here, \(\sigma_{\min}(V)\) denotes the smallest singular value of the query-difference matrix. Equation~\eqref{eq:sigma} shows that recovery accuracy depends jointly on score error and the conditioning of the query embeddings. Homomorphic-computation error alone therefore does not prevent template inference.

To validate the attack path described by Theorem 1, we simulate template recovery under FLEA. The corrupted client belongs to user \(X\) and knows the feature extractor \(f_{\theta}(\cdot)\), its own plaintext query features \(e_{i,a}^{Q}\), and the continuous scores \(S_{i}^{Y}\) returned by the server. We evaluate template recovery for every source--target identity pair,
treating each \(X\!\to\!Y\) pair as one attack instance. Each query contains three consecutive EEG windows. 
For request $i$, let $S_i^Y$ denote the observed normalized score and define $a_i=\sum_{a=0}^{2}\|\mathbf{e}_{i,a}^Q\|_2^2$. From the three-window score definition, subtracting the score for reference request~0 eliminates the fixed threshold and gives:
\begin{equation}
2(\bar{\mathbf{e}}_i^Q-\bar{\mathbf{e}}_0^Q)^\top\mathbf{t}_Y
=
\frac{a_i-a_0}{3}-(S_i^Y-S_0^Y).
\label{eq:empirical-dltr}
\end{equation}
Queries are added sequentially to form the template-recovery system. We record the rank and singular values of the resulting difference matrix and the template-recovery error. Table~\ref{tab:flea-recovery} reports the results when the system first reaches full rank and when all available queries are used. The relative recovery error is:
\begin{equation}
    E_{\mathrm{rel}} =
    \frac{\|\hat{\mathbf{t}}_Y-\mathbf{t}_Y\|_2}
         {\|\mathbf{t}_Y\|_2}.
    \label{}
\end{equation}

\begin{table}[t]
\caption{Template-Recovery Results under FLEA}
\label{tab:flea-recovery}
\centering
\footnotesize
\begin{tabular*}{\columnwidth}{@{\extracolsep{\fill}}lcc@{}}
\toprule
\textbf{Measure} & \textbf{HGD} & \textbf{BCICIV-2a} \\
\midrule
Attack instances & 91 & 40 \\
Queries per attack & 158 & 286 \\
Queries to rank 32 (successes) & 33 (91/91) & 33 (40/40) \\
\midrule
\multicolumn{3}{l}{Relative error at 33 queries} \\
\quad Mean & $1.850\times10^{-3}$ & $4.080\times10^{-4}$ \\
\quad Maximum & $1.572\times10^{-2}$ & $3.101\times10^{-3}$ \\
\multicolumn{3}{l}{Relative error using all queries} \\
\quad Mean & $5.523\times10^{-5}$ & $1.703\times10^{-5}$ \\
\quad Maximum & $1.048\times10^{-4}$ & $4.109\times10^{-5}$ \\
Minimum cosine similarity at 33 queries
& 99.9876\% & 99.9995\% \\
\bottomrule
\end{tabular*}
\end{table}

All \(X\!\to\!Y\) pairs reached rank 32 after the 33rd request, consistent with the $d+1$ query condition in Theorem 1. Using only the 33 requests required to reach full rank, the mean relative errors on HGD and BCICIV-2a were $ 1.850\times {10}^{-3} $ and $ 4.080\times {10}^{-4} $, respectively. Using the complete query pool for X and solving the resulting overdetermined system by least squares reduced the mean errors to $ 5.523\times {10}^{-5}$ and $ 1.703\times {10}^{-5}$. At 33 queries, the maximum error exceeded the mean; the maximum condition numbers of the HGD and BCICIV-2a difference matrices were approximately $ 8.71\times {10}^{5}$ and $ 5.85\times {10}^{4}$, respectively. Thus, recovery accuracy at the minimum full-rank query count remains sensitive to matrix conditioning, while additional queries improve the estimate. This experiment validates the template-recovery risk under FLEA and motivates the proposed end-to-end encrypted architecture.

\textbf{Proposition 1 (Impediment to Template Recovery without Known Features)}

Suppose the TopoCryptoEEG feature-extractor parameters are held only by the server and the client cannot directly obtain plaintext query embeddings. The known coefficient matrix V required by Theorem 1 is then absent from the client’s directly observable protocol view, so the direct linear solution used against FLEA cannot be applied in the same way. Furthermore, assume that the permitted server-model family is closed under a global sign change of the output embedding and that the client receives only continuous authentication scores. Under these assumptions, the absolute coordinates of the registered template cannot be uniquely determined from the scores alone.

\textbf{Proof:}
By \eqref{eq:matrix v&delta}, direct algebraic recovery requires the client to know:
\begin{equation}
    \label{eq:matrix minors}
    V=\begin{bmatrix}{\left({\overline{e}}_{1}^{Q}-{\overline{e}}_{0}^{Q}\right)}^{\top } \\ \vdots  \\ {\left({\overline{e}}_{d}^{Q}-{\overline{e}}_{0}^{Q}\right)}^{\top }\end{bmatrix}.
\end{equation}

In the conventional scheme, the client’s local model generates these query embeddings $e_{i,a}^{Q}$, so the client can construct $V$.

In contrast, under TopoCryptoEEG: $e_{i,a}^{Q}=f_{\theta }\left(x_{i,a}\right),$
Here, $\theta$ is a server-private parameter and the protocol does not release the embedding $e_{i,a}^{Q}$. Although the client knows $x_{i,a}$ and $S_{u,i}$, it cannot directly compute the actual feature differences in \eqref{eq:matrix minors}.Now consider two possible server secret states:
$ (\theta, e^T)$ , $ (\theta', e^{T'}) $
where:
\begin{equation}
    \label{eq:secret states}
    f_{\theta' }\left(x\right)=-f_{\theta}\left(x\right),\quad e^{T'}=-e^{T}.
\end{equation}
Because TopoFuse-MLP produces its output through a linear fusion layer (see Section IV-B), the feature transformation in \eqref{eq:secret states} can be implemented by negating both the weights and bias of the final fusion layer, which remains within the same inference-model family.

For any valid request $X_{i}$, the scores produced by the two secret states satisfy:
\begin{equation}
    \label{}
    \begin{aligned}{S'}_{u,i}=\sum_{a=0}^{q-1} {\left\lVert -f_{\theta }\left(x_{i,a}\right)+e^{T}\right\lVert }_{2}^{2}-q\tau  \\ =\sum_{a=0}^{q-1} {\left\lVert f_{\theta }\left(x_{i,a}\right)-e^{T}\right\lVert }_{2}^{2}-q\tau  =S_{u,i}.\end{aligned} 
\end{equation}

Consequently, the client receives exactly the same sequence of continuous scores under both secret states. This equivalence also holds for adaptive queries: because the response histories are identical, the attacker’s subsequent inputs have the same conditional distribution.

However, the two states have templates $ e^{T}$ and $ -e^{T}$ and query-embedding matrices V and -V, respectively. When $e^{T}\neq 0$, the template coordinates differ.

Therefore, from this score interface alone, the client cannot uniquely determine the absolute coordinates of the registered template across these equivalent secret states, nor can it uniquely determine the true query-feature matrix in Theorem 1 from its protocol view. This proves the proposition.

Thus, the known-feature linear recovery attack cannot be directly applied under the stated assumptions; other score-based inference attacks are outside the scope of this analysis.

\subsection{System Overview}

Figure 2 overviews TopoCryptoEEG, which consists of a client and a server. The client acquires the signal, performs public preprocessing, constructs the topology-aware input, and encrypts it. The server retains the identity feature extractor and performs enrollment-template generation and online encrypted authentication.

\begin{figure*}[!t]
\centering
\includegraphics[width=\textwidth]{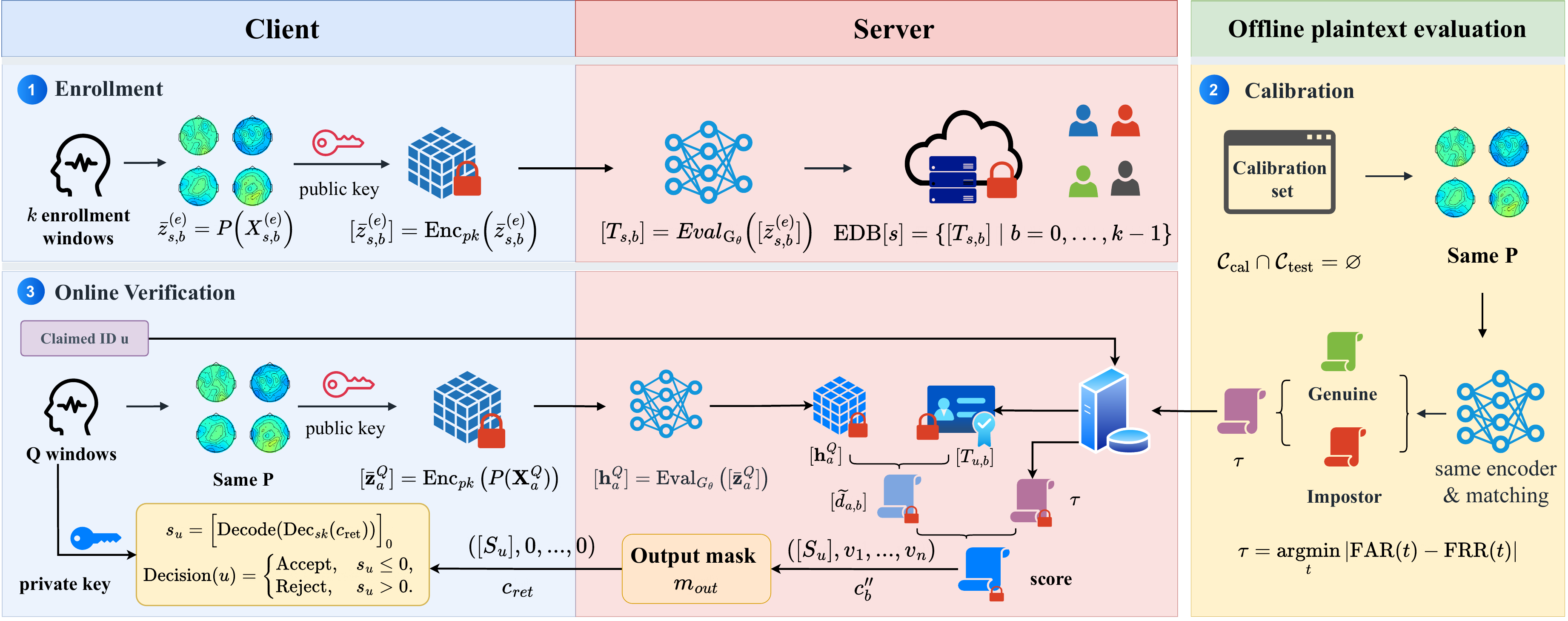}
\caption{Overview of TopoCryptoEEG.}
\label{fig_2}
\end{figure*}
The authentication protocol has three logical phases: offline training and enrollment, offline threshold calibration, and online authentication. Training, enrollment, and threshold calibration are completed before deployment and are excluded from per-request online latency. Online authentication reuses the fixed model, threshold, and registration templates and processes the current encrypted topology-aware input.

During offline enrollment, the client generates a homomorphic-encryption key set for a key domain, covering identity records and authentication requests in that domain. The secret key $sk$ remains with the client; the public key $pk$ and evaluation keys are supplied to the server. For the $b$-th enrollment window of identity $s$, the client constructs and encrypts the topology-patch input. The server performs local-branch encoding in the ciphertext domain to obtain the encrypted registration template:
\begin{equation}
    \label{}
    \left[T_{s,b}\right]=\operatorname{Eval}_{G_{\theta }}{\left(\left[{\mathbf{z}}_{s,b}^{T}\right]\right)},
\end{equation}
Here, $G_{\theta }$ denotes the local-branch encoder, and the template’s logical representation is:
\begin{equation}
    \label{}
    T_{s,b}={\mathbf{h}}_{s,b,0}^{T,\left(2\right)}\Vert \cdots \Vert {\mathbf{h}}_{s,b,P-1}^{T,\left(2\right)}.
\end{equation}
The server caches the ciphertexts of the branch representations and associates them with the identity and enrollment-window indices. The prescribed enrollment protocol does not require either party to decrypt these templates.

Offline threshold calibration: In an independent plaintext evaluation procedure, the trained model is applied to the calibration set to generate representations. Distances are computed for genuine and impostor claims, and the threshold is selected according to the equal-error-rate criterion:
\begin{equation}
    \label{}
    \tau =\arg{\min_{t}{\left|\operatorname{FAR}_{\mathrm{cal}}{\left(t\right)}-\operatorname{FRR}_{\mathrm{cal}}{\left(t\right)}\right|}}.
\end{equation}

The calibration and test sets are disjoint. The selected threshold is included as a plaintext constant in the encrypted test claim and participates in online ciphertext-domain score computation; threshold calibration is not performed during encrypted testing.

We summarize the online authentication phase in Algorithm \ref{alg:online-auth}.

\section{Proposed Methods}
\subsection{Multi-Window Topology-Aware Encoding}
EEG signals exhibit temporal variation and cross-channel spatial structure. Topographic mapping preserves electrode relationships in a two-dimensional representation, as explored in prior work~\cite{ref37}. To support encrypted computation while controlling input size, MWTE maps multiband features to a fixed grid, then applies pooling and nonoverlapping patch partitioning. The $q$ windows in each request are encoded separately, and their matching distances are aggregated (Fig.~3).

\begin{figure*}[!t]
\centering
\includegraphics[width=\textwidth]{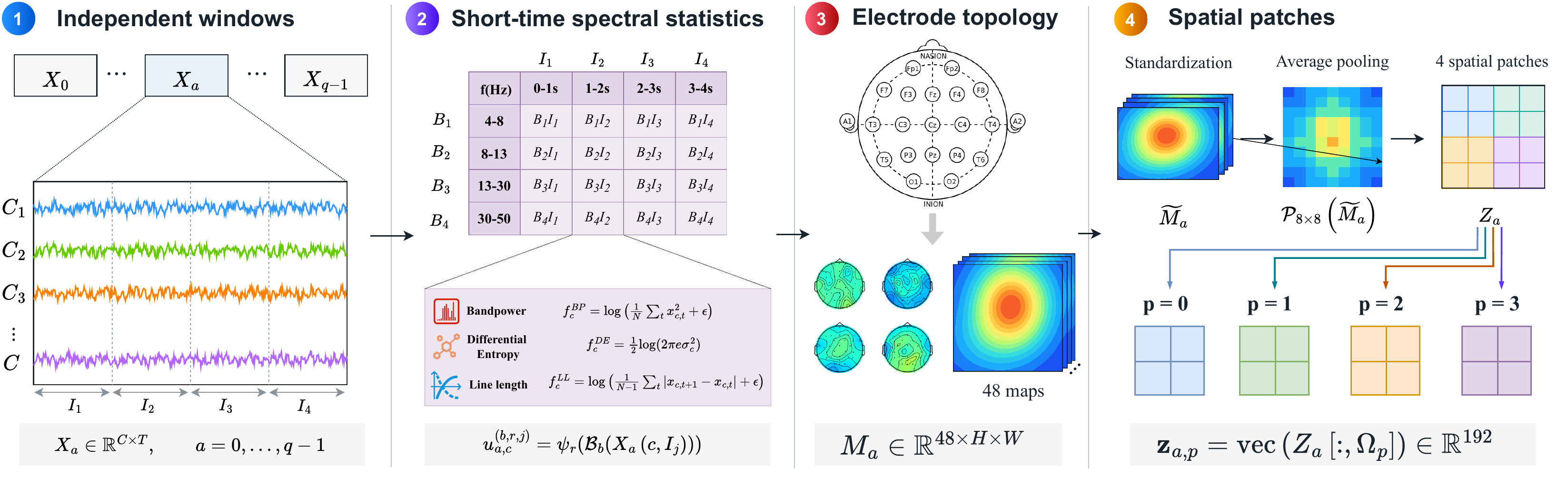}
\caption{Multi-window topology-aware encoding.}
\label{fig_3}
\end{figure*}
Suppose an authentication request contains q EEG windows. The $a$-th window is denoted by:
$X_{a}\in {\mathbb{R}}^{C\times T},\quad \quad a=0,\ldots ,q-1,$
Here, $C$ is the number of EEG channels and $T$ is the number of samples in a window. For each window, the client extracts bandpower, differential entropy, and line length.

To capture short-term dynamics in the topographic representation, each window is further divided into $J$ segments, and the band features are computed for each segment. Let $u_{a,c}^{\left(b,r,j\right)}={\psi }_{r}\! \left({\mathcal{B}}_{b}\! \left(X_{a}\left(c,I_{j}\right)\right)\right)$ denote the scalar feature for window $a$, channel $c$, frequency band $b$, statistic $r$, and segment $j$: Here, ${\mathcal{B}}_{b}\left(\cdot \right)$ denotes the transform for frequency band b, ${\psi }_{r}\left(\cdot \right)$ denotes the statistical-feature extractor, and $I_{j}$ denotes the $j$-th time segment. With four frequency bands, three statistics, and four temporal segments, this yields 48 channel-level features.

The client then maps each channel-level feature to a two-dimensional topographic map according to the scalp-plane coordinates of the EEG electrodes. Let: $\varepsilon ={\left\{l_{c}\right\}}_{c=1}^{C}$ denote the electrode-location set, and let $\mathcal{I}\left(\cdot ,\varepsilon \right)$ denote the spatial interpolation operator determined by electrode topology. The topographic map for window $a$ and feature indices $b$, $r$, and $j$ is:
\begin{equation}
    \label{}
    M_{a}^{\left(b,r,j\right)}=\mathcal{I}\! \left(u_{a,1}^{\left(b,r,j\right)},\ldots ,u_{a,C}^{\left(b,r,j\right)};\varepsilon \right)\in {\mathbb{R}}^{H\times W}.
\end{equation}

Stacking all topographic maps along the feature dimension gives the window-level topology-aware representation:
\begin{equation}
    \label{}
    M_{a}=\operatorname{Stack}{\left(M_{a}^{\left(b,r,j\right)}\right)}\in {\mathbb{R}}^{48\times H\times W}.
\end{equation}

To control the input size for encrypted encoding, the client standardizes each topographic map and applies spatial average pooling. Let ${\tilde{M}}_{a}$ denote the standardized window-level topographic representation. With $H=W=32$, average pooling uses a kernel and stride of 8, yielding:
\begin{equation}
    \label{}
    Z_{a}={\mathcal{P}}_{8\times 8}\left({\tilde{M}}_{a}\right)\in {\mathbb{R}}^{48\times 4\times 4}.
\end{equation}

The pooled spatial grid is then partitioned into $P=4$ nonoverlapping 2 × 2 topology-aware patches. Let ${\Omega }_{p}$ denote the spatial indices of patch p in the pooled grid. Its input is:
\begin{equation}
    \label{}
    {\mathbf{z}}_{a,p}=\operatorname{vec}{\! }\left(Z_{a}\left[:,{\Omega }_{p}\right]\right)\in {\mathbb{R}}^{d_{0}}.
\end{equation}

Among them, $d_{0}=192$. Each patch retains the features of its spatial region, and a fixed patch order preserves correspondence across query and enrollment windows for branch encoding and ciphertext packing.

\subsection{Topographic Branch Model with Linear Fusion}

To address the tension between the representational capacity of deep EEG-decoding models and the limited multiplicative depth available for homomorphic inference, we construct a lightweight topographic branch model with linear fusion to extract identity representations from the topology-aware patches. Low-degree polynomial activations and shallow local branches offer complementary ways to control circuit depth and enable blockwise inference \cite{ref10,ref11}. Each local branch therefore contains two linear layers, each followed by a square activation. The four branch outputs are combined by linear fusion to produce a window-level identity embedding. Local nonlinearities extract regional patterns, while linear fusion integrates information across regions without adding further nonlinear layers. Query and enrollment windows use the same model parameters. Because different scalp regions may contain different identity-related patterns, each topology-aware patch is assigned an independent encoder. Branch p first extracts a local representation from its corresponding region:
    \begin{align}
	{\mathbf{h}}_{a,p}^{\left(1\right)}=\phi \left(W_{1,p}{\mathbf{z}}_{a,p}+{\mathbf{b}}_{1,p}\right),\ \ W_{1,p}\in {\mathbb{R}}^{32\times 192},\\{\mathbf{h}}_{a,p}^{\left(2\right)}=\phi \left(W_{2,p}{\mathbf{h}}_{a,p}^{\left(1\right)}+{\mathbf{b}}_{2,p}\right),\ \ W_{2,p}\in {\mathbb{R}}^{32\times 32},
\end{align}

Here, $\phi \left(\mathbf{x}\right)=\mathbf{x}\odot \mathbf{x}$ is the elementwise square activation.
The outputs of the four branches are concatenated in spatial patch order to form the joint local representation:
\begin{equation}
    \label{}
    {\mathbf{h}}_{a}={\mathbf{h}}_{a,0}^{\left(2\right)}\Vert \cdots \Vert {\mathbf{h}}_{a,P-1}^{\left(2\right)}\in {\mathbb{R}}^{Pd_{e}}.
\end{equation}

Partition the fusion matrix by patch as follows:
\begin{equation}
    \label{}
    W_{f}=\left[W_{f,0}\ \cdots \ W_{f,P-1}\right],\ \ W_{f,p}\in {\mathbb{R}}^{D\times d_{e}}.
\end{equation}

The window-level identity embedding is obtained by linear fusion:
\begin{equation}
    \label{}
    {\mathbf{e}}_{a}=W_{f}{\mathbf{h}}_{a}+{\mathbf{b}}_{f}=\sum_{p=0}^{P-1} W_{f,p}{\mathbf{h}}_{a,p}^{\left(2\right)}+{\mathbf{b}}_{f}.
\end{equation}

We set $d_{e}=D=32$; the fusion layer therefore maps the 128-dimensional joint local representation to a 32-dimensional identity embedding. 

After constructing the topology-aware patches, the client encrypts the input representation:
\begin{equation}
    \label{}
    \left[{\mathbf{z}}_{a,p}\right]={\mathrm{Enc}}_{pk}\left({\mathbf{z}}_{a,p}\right).
\end{equation}

The server executes the branch encoders and fusion in the ciphertext domain:
    \begin{align}
    \label{}
    \left[{\mathbf{h}}_{a,p}^{\left(1\right)}\right]=\phi \left(W_{1,p}\left[{\mathbf{z}}_{a,p}\right]+{\mathbf{b}}_{1,p}\right),\\
    \left[{\mathbf{h}}_{a,p}^{\left(2\right)}\right]=\phi \left(W_{2,p}\left[{\mathbf{h}}_{a,p}^{\left(1\right)}\right]+{\mathbf{b}}_{2,p}\right),\\
    \left[{\mathbf{e}}_{a}\right]=W_{f}\left({\Vert }_{p=0}^{P-1}\left[{\mathbf{h}}_{a,p}^{\left(2\right)}\right]\right)+{\mathbf{b}}_{f}.
\end{align}

The server computes the fused embeddings in the ciphertext domain and evaluates their squared distances to obtain the authentication score:
\begin{equation}
    \label{}
    d_{a,b}={\left\lVert {\mathbf{e}}_{a}^{Q}-{\mathbf{e}}_{b}^{T}\right\lVert }_{2}^{2},\quad \quad S=\frac{1}{qk}\sum_{a,b} d_{a,b}-\tau .
\end{equation}

\subsection{Encrypted-Computation Optimizations}

\subsubsection{\textbf{Query–Patch Coordinate Packing}}

Prior homomorphic neural-network studies use slot packing and layout optimization to process tensor elements in parallel, reducing rotations or data replication in linear transformations \cite{ref17,ref18,ref19,ref20,ref21}. Layouts must match tensor shapes and operator access patterns. Query windows and topology-aware patches have explicit two-dimensional indices; processing them independently as ordinary contiguous vectors prevents repeated evaluations across windows and patches from being shared. We therefore map both indices to unified query–patch lanes, aligning the same logical coordinate across lanes to share ciphertext evaluation while preserving the correspondence of each local input. For an authentication request containing $q$ query windows and $P$ topology-aware patches, the lane index of patch $p$ in window a is: $\mathrm{\ell }\left(a,p\right)=aP+p.$
Let \(L \geq qP\) denote the slot interval between adjacent block coordinates. The client first constructs a packed query vector $x_{Q}$ whose valid input region satisfies:
\begin{equation}
    \label{}
    {x}_{Q}\left[Lf+\mathrm{\ell }\left(a,p\right)\right]={\mathbf{z}}_{a,p}^{Q}\left[f\right],\ \ f=0,\ldots ,d_{0}-1,
\end{equation}

The client then encrypts $\left[{\mathbf{x}}_{Q}\right]$. In the current setting $q=3$ and $P=4$, there are 12 valid lanes; we use $L=16$ and set the remaining four lanes to zero.
The $b$-th enrollment window uses the same layout:
\begin{equation}
    \label{}
    {x}_{T,b}\left[Lf+\mathrm{\ell }\left(a,p\right)\right]={\mathbf{z}}_{b,p}^{T}\left[f\right].
\end{equation}

\subsubsection{\textbf{Lane-Preserving BSGS Evaluation}}

BSGS decomposes diagonal-form linear transformations into baby-step and giant-step rotations to reduce ciphertext rotations \cite{ref22}. Its rotation indices depend on the ciphertext-slot layout. Under our multi-lane layout, directly applying rotations defined for a logical vector may cross lane boundaries and disrupt query-window and patch correspondence. We therefore scale logical rotations by the lane interval when mapping them to physical rotations, allowing BSGS to share rotation computation while keeping lanes independent. Since adjacent logical block coordinates are $L$ physical slots apart, a logical feature rotation $r$ maps to the physical rotation $Lr$:
\(r \mapsto Lr\) .
For any slot coordinate $Lf+\mathrm{\ell }\left(a,p\right)$:
\begin{equation}
    \label{}
    Lf+\mathrm{\ell }\left(a,p\right)+Lr=L\left(f+r\right)+\mathrm{\ell }\left(a,p\right).
\end{equation}

Represent the linear transformation in diagonal form and write the diagonal index as: $r=i+\beta j$,
where $\beta$ is the baby-step parameter. The lane-aware BSGS linear transformation is then:
\begin{equation}
    \label{}
    {\mathcal{L}}_{W}\left(X\right)=\sum_{j} {\rho }_{L\beta j}\left(\sum_{i} {\rho }_{-L\beta j}\left({\mathbf{D}}_{r}\right)\odot {\rho }_{Li}\left(X\right)\right)+B,
\end{equation}

Here, ${\rho }_{t}$ denotes a left rotation by t physical slots, while ${\mathbf{D}}_{r}$ and $B$ are the plaintext diagonal-weight vector and bias, respectively. For a linear layer with input dimension $n_{\mathrm{in}}$, the diagonal weights at a valid output coordinate f satisfy:
\begin{equation}
    \label{}
    {\mathbf{D}}_{r}\left[Lf+\mathrm{\ell }\left(a,p\right)\right]=W_{p}\! \left[f,\,\left(f+r\right)\bmod n_{\mathrm{in}}\right],
\end{equation}

Here, $W_{p}$ is the weight matrix for branch $p$. Weights for the same patch are reused across query-window lanes, while lanes belonging to different patches contain their corresponding branch weights. A shared baby-step rotation can therefore serve multiple windows and branches with different parameters. The auxiliary coordinate region is filled with weights and biases according to the periodic extension rule of the corresponding linear layer. Fig. 4 illustrates the design.
\begin{figure*}[!t]
\centering
\includegraphics[width=\textwidth]{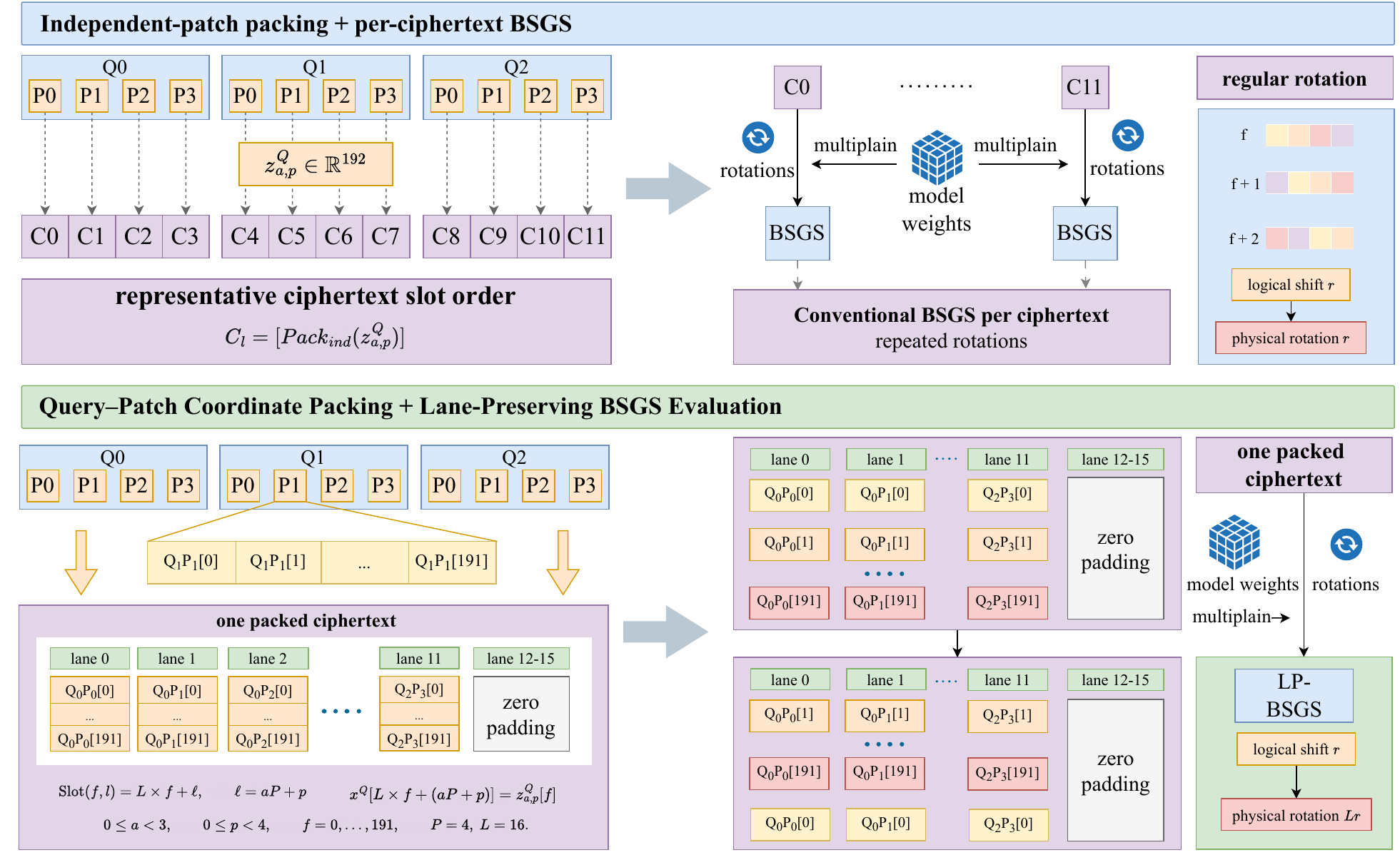}
\caption{Query–patch coordinate packing and lane-preserving BSGS evaluation. A lane is determined jointly by the window and patch; rotations along the block-coordinate dimension are shared across lanes.}
\label{fig_4}
\end{figure*} 

\subsubsection{\textbf{Slot Alignment for the Fusion Layer}}

After the two local-branch encoding layers, the query and enrollment-template outputs have the same slot layout. For query window $a$, template $b$, and patch $p$, the server computes the branch difference
\begin{equation}
    \label{}
    \delta \mathbf{h}_{a,b,p}=\mathbf{h}_{a,p}^{Q,(2)}-\mathbf{h}_{b,p}^{T,(2)}.
\end{equation}

Since the query and template use the same fusion parameters, their fused-embedding difference is
\begin{equation}
    \label{}
    \boldsymbol{\Delta}_{a,b}
    =
    \mathbf{e}_{a}^{Q}-\mathbf{e}_{b}^{T}
    =
    \sum_{p=0}^{P-1} \mathbf{W}_{f,p}\delta\mathbf{h}_{a,b,p}.
\end{equation}

The fusion bias cancels in this difference. The server therefore projects each branch difference independently and packs the result into $\mathbf{g}_b$:
\begin{equation}
    \label{}
    \mathbf{g}_b[Ld+aP+p]
    =
    \left(\mathbf{W}_{f,p}\delta\mathbf{h}_{a,b,p}\right)[d],
    \quad d=0,\ldots,D-1.
\end{equation}

Using the left-rotation convention, the server aligns the patch projections to the first patch lane of each query window:
\begin{equation}
    \label{}
    \mathbf{v}_b
    =
    \mathbf{m}_f \odot
    \sum_{p=0}^{P-1}\rho^{p}(\mathbf{g}_b),
\end{equation}

where $\mathbf{m}_f$ retains slots $Ld+aP$. Thus,
\begin{equation}
    \label{}
    \mathbf{v}_b[Ld+aP]=\Delta_{a,b}[d].
\end{equation}

\subsubsection{\textbf{Ciphertext Distance Computation and Reduction}}

For each enrollment template, the server computes the squared embedding distance for every query window, sums the distances across windows and templates, and subtracts the threshold. Algorithm \ref{alg:online-auth} summarizes the online ciphertext-domain computation. For $D=32$, feature reduction uses rotations by $L,2L,4L,8L,$ and $16L$. In the current $q=3$, $P=4$ setting, the query-window distances occupy slots $0,4,$ and $8$ and are reduced using rotations by $4$ and $8$ slots.

\begin{algorithm}[t]
\caption{Online Encrypted Brainprint Authentication}
\label{alg:online-auth}
\begin{algorithmic}[1]
\REQUIRE Encrypted QPCP query $[ \mathbf{x}^{Q} ]$;
cached enrollment branch outputs
$\{[\mathbf{h}_{b,p}^{T,(2)}]\}_{b=0,p=0}^{k-1,P-1}$;
fusion weights $\{\mathbf{W}_{f,p}\}_{p=0}^{P-1}$;
threshold $\tau$; $q=3$, $k=1$, $P=4$, $D=32$, lane interval $L=16$
\ENSURE Encrypted threshold-adjusted score $[ c_{\mathrm{ret}} ]$
\STATE Evaluate the query branches on $[ \mathbf{x}^{Q} ]$
using LP-BSGS to obtain
$\{[\mathbf{h}_{a,p}^{Q,(2)}]\}_{a=0,p=0}^{q-1,P-1}$
\FOR{$b=0$ to $k-1$}
    \FOR{$a=0$ to $q-1$}
        \FOR{$p=0$ to $P-1$}
            \STATE $[\delta\mathbf{h}_{a,b,p}]
            \gets [\mathbf{h}_{a,p}^{Q,(2)}]
            -[\mathbf{h}_{b,p}^{T,(2)}]$
        \ENDFOR
    \ENDFOR
    \STATE Compute $[\mathbf{g}_b]$ by applying the patch-specific projections $\mathbf{W}_{f,p}$ to all $[\delta\mathbf{h}_{a,b,p}]$ and packing the outputs at slots $Ld+aP+p$ using LP-BSGS
    \STATE $[\mathbf{v}_{b}]
    \gets \mathbf{m}_{f}\odot
    \sum_{p=0}^{P-1}\rho^{p}([\mathbf{g}_{b}])$
    \STATE $[\mathbf{c}_{b}^{(0)}]
    \gets [\mathbf{v}_{b}]
    \odot [\mathbf{v}_{b}]$
    \FOR{$t=0$ to $\log_{2}D-1$}
        \STATE $[\mathbf{c}_{b}^{(t+1)}]
        \gets [\mathbf{c}_{b}^{(t)}]
        + \rho^{2^{t}L}([\mathbf{c}_{b}^{(t)}])$
    \ENDFOR
    \STATE $[\mathbf{s}_{b}]
    \gets [\mathbf{c}_{b}^{(\log_{2}D)}]$
    \IF{$q=3$}
        \STATE $[\mathbf{s}_{b}]
        \gets [\mathbf{s}_{b}]
        + \rho^{4}([\mathbf{s}_{b}])$
        \STATE $[\mathbf{s}_{b}]
        \gets [\mathbf{s}_{b}]
        + \rho^{8}([\mathbf{s}_{b}])$
    \ENDIF
\ENDFOR
\STATE $[ c_{\mathrm{ret}} ]
\gets \mathbf{m}_{\mathrm{out}}\odot
\left(\sum_{b=0}^{k-1}[\mathbf{s}_{b}]
-qk\tau\right)$
\STATE Return $[ c_{\mathrm{ret}} ]$; the client decrypts
slot $0$ and accepts iff its value is negative
\end{algorithmic}
\end{algorithm}

\section{Experiments}
This section first evaluates template-recovery attacks against FLEA and then assesses the authentication performance and end-to-end encrypted-computation efficiency of TopoCryptoEEG. We describe the platform, datasets, evaluation metrics, and experimental details before presenting a comprehensive evaluation.

Because no end-to-end encrypted brainprint-authentication scheme is currently available for direct comparison, we select two baselines from related areas to evaluate system performance: CryptoFace\cite{ref11}, a representative end-to-end encrypted biometric-recognition system, and EER \cite{ref9}, an encrypted EEG emotion-recognition scheme. Both baselines accept single-window samples, so we compare them with the single-window configuration of TopoCryptoEEG and additionally report TopoCryptoEEG results for three windows.

To evaluate the lightweight feature extractor TopoFuse-MLP, we compare it with three advanced brainprint-authentication models—BrainprintNet \cite{ref38}, ADN \cite{ref39}, and DARN \cite{ref40}—under the same plaintext authentication protocol and a three-window configuration, retaining each model’s architecture and nonlinear operators. We also assess plaintext–ciphertext decision consistency and score error, and implement Zhang et al.’s feature-level encrypted method (FLEA) using the same feature extractor to isolate the encrypted-computation cost introduced by TopoCryptoEEG.

Finally, we conduct ablation studies of MWTE and the encrypted-computation optimizations. Four MWTE configurations examine the effects of multi-window aggregation, spatial-topology mapping, and statistical-feature representations. Five encrypted-computation configurations use operator counts and measured latency to assess the individual contributions of QPCP, LP-BSGS, and GPU execution.

\subsection{Experimental Setup}

\subsubsection{\textbf{Test Platform}}
TopoCryptoEEG uses Microsoft.Research.SEALNet 4.1.1 for CPU inference and HEonGPU 1.1 for GPU inference \cite{ref41,ref42}. Experiments run on an Intel Core i9-14900KF CPU and an NVIDIA GeForce RTX 4090D GPU with 24~GB of memory. We use leveled CKKS with polynomial modulus degree $N=16{,}384$ (8,192 complex slots), coefficient-modulus bit sizes $(60,40,40,40,40,40,40,40,60)$, and initial scale $2^{40}$. These parameters pass Microsoft SEAL's TC128 security validation \cite{ref34,ref43}.

\subsubsection{\textbf{Datasets}}
We use HGD (500~Hz, 44 channels)~\cite{ref44} and BCICIV-2a (250~Hz, 22 channels)~\cite{ref45}; dataset sizes and partitions are listed in Table~\ref{tab:dataset_partition}. Each dataset is trained on Run~1 windows from all subjects. Enrollment templates are selected from the same Run~1 training pool after model fitting, whereas calibration and test queries are drawn from Run~2. Calibration and test subjects are disjoint, but both groups are represented during model training. The protocol therefore evaluates seen-identity cross-session authentication. EEG is segmented into 4-s windows with a 4-s stride.

\begin{table}[!t]
\centering
\caption{Dataset size and partitioning}
\label{tab:dataset_partition}
\begin{tabular}{lcc}
\toprule
\textbf{Metric} & \textbf{HGD} & \textbf{BCICIV-2a} \\
\midrule
Total subjects          & 14   & 9    \\
Calibration subjects    & 7    & 4    \\
Test subjects           & 7    & 5    \\
Training windows        & 4480 & 2592 \\
Calibration windows     & 1120 & 1152 \\
Test windows            & 1120 & 1440 \\
Test claims per group   & 4424 & 5720 \\
\bottomrule
\end{tabular}
\end{table}

\subsubsection{\textbf{Evaluation Metrics}}
Authentication performance is measured by ROC-AUC, EER, balanced accuracy (BA), and TAR at FAR levels of 1\% and 5\%, where $\mathrm{BA}=(\mathrm{TAR}+\mathrm{TNR})/2$. Plaintext--ciphertext consistency is assessed by decision agreement, mean and maximum absolute score errors, and minimum decision margin. Scores use the same threshold and scaling. Efficiency is measured by per-claim encrypted-computation latency, RAM, and GPU memory. Operator counts include input encryption, MultiplyPlain, Rotate, and Square.

\subsubsection{\textbf{Experimental Details}}

\textbf{Evaluation protocol:} We evaluate $q=1$ and $q=3$ using the same data splits, enrollment-template indices, and query lists across methods. For each identity, the median enrollment window in temporal order is used as the single template ($k=1$). Each multi-window query contains three consecutive windows from the same subject and session; window distances are averaged. Encrypted baselines use $q=1$, and TopoCryptoEEG is reported for both settings. Plaintext model comparisons use $q=3$, with one genuine and three impostor claims per query start. Results are averaged over 15 combinations of five training seeds and three claim seeds, and reported as mean $\pm$ standard deviation.

\textbf{Training and baselines:} TopoFuse-MLP is trained for 20 epochs using AdamW (learning rate $10^{-3}$, weight decay $10^{-4}$, batch size 32) with identity-classification and contrastive losses; its classification head is removed at inference. For CryptoFace, $32\times32$ topographic maps are resized to $64\times64$, with a reduced rotation-key set under original security parameters. The EER baseline uses a dual-view MLPLogistic model. BrainprintNet, ADN, and DARN use raw EEG resampled to 128~Hz, while TopoFuse-MLP uses topology-aware inputs.

\textbf{Timing and ablations:} Plaintext--ciphertext comparisons use 40 fixed requests (20 genuine and 20 impostor). Each timing run has five warm-up requests and measures 40 requests; latency is repeated five times. Online timing starts after query-patch construction and excludes signal acquisition, transmission, and offline enrollment. MWTE ablations use the same 15 seed combinations; configurations A and B are tuned independently (learning rate $5\times10^{-4}$, 30 epochs). Encrypted-computation ablations fix the model, CKKS parameters, and request set across H0--H4; the two BSGS transformations use baby-step parameters 16 and 8.

\subsection{Evaluation Results}

Tables \ref{tab:hgd_comparison} and \ref{tab:bciciv_comparison} compare TopoCryptoEEG with encrypted baselines on the two datasets.
\begin{table*}[!t]
\centering
\caption{Comparison of encrypted schemes on the HGD dataset.}
\label{tab:hgd_comparison}
\small
\setlength{\tabcolsep}{6pt} 
\begin{tabular}{lccccccc}
\toprule
\textbf{Method} & \makecell{\textbf{AUC} \\ (\%)} & \makecell{\textbf{EER} \\ (\%)} & \makecell{\textbf{BA} \\ (\%)} & \makecell{\textbf{TAR@} \\ \textbf{FAR=1\%} (\%)} & \makecell{\textbf{TAR@} \\ \textbf{FAR=5\%} (\%)} & \makecell{\textbf{Latency} \\ per claim (s)} & \makecell{\textbf{RAM} \\ (GiB)} \\
\midrule
CryptoFace \cite{ref11} & 98.04 \(\pm\) 0.41 & 4.75 \(\pm\) 0.83 & 93.94 \(\pm\) 1.93 & 86.15 \(\pm\) 6.67 & 95.26 \(\pm\) 1.30 & 1416.45 & 47 \\
EER \cite{ref9} & 95.81 \(\pm\) 0.39 & 10.35 \(\pm\) 0.42 & 90.22 \(\pm\) 0.28 & 36.57 \(\pm\) 6.49 & 76.18 \(\pm\) 2.96 & 27.00 & 16.18 \\
\makecell[l]{\textbf{TopoCryptoEEG} \\ \textbf{(ours, \(q=1\))}} & \textbf{99.50 \(\pm\) 0.16} & \textbf{3.15 \(\pm\) 0.56} & \textbf{96.63 \(\pm\) 0.70} & \textbf{91.61 \(\pm\) 1.86} & \textbf{97.94 \(\pm\) 0.93} & \textbf{1.616} & \textbf{13.62} \\
\makecell[l]{\textbf{TopoCryptoEEG} \\ \textbf{(ours, \(q=3\))}} & \textbf{99.93 \(\pm\) 0.03} & \textbf{1.21 \(\pm\) 0.28} & \textbf{98.51 \(\pm\) 0.52} & \textbf{98.28 \(\pm\) 1.04} & \textbf{99.83 \(\pm\) 0.19} & \textbf{1.578} & \textbf{14.03} \\
\bottomrule
\end{tabular}
\end{table*}

\begin{table*}[!t]
\centering
\caption{Comparison of encrypted schemes on the BCICIV-2a dataset.}
\label{tab:bciciv_comparison}
\small
\setlength{\tabcolsep}{6pt}
\begin{tabular}{lccccccc}
\toprule
\textbf{Method} & \makecell{\textbf{AUC} \\ (\%)} & \makecell{\textbf{EER} \\ (\%)} & \makecell{\textbf{BA} \\ (\%)} & \makecell{\textbf{TAR@} \\ \textbf{FAR=1\%} (\%)} & \makecell{\textbf{TAR@} \\ \textbf{FAR=5\%} (\%)} & \makecell{\textbf{Latency} \\ per claim (s)} & \makecell{\textbf{RAM} \\ (GiB)} \\
\midrule
CryptoFace \cite{ref11} & 69.82 \(\pm\) 1.24 & 33.48 \(\pm\) 1.63 & 62.26 \(\pm\) 1.83 & 19.60 \(\pm\) 3.51 & 36.89 \(\pm\) 3.70 & 1451.9 & 56 \\
EER \cite{ref9} & 72.02 \(\pm\) 0.66 & 34.46 \(\pm\) 0.35 & 65.18 \(\pm\) 0.49 & 7.10 \(\pm\) 0.26 & 23.72 \(\pm\) 0.77 & 28.58 & 16.26 \\
\makecell[l]{\textbf{TopoCryptoEEG} \\ \textbf{(ours, \(q=1\))}} & \textbf{96.21 \(\pm\) 0.11} & \textbf{9.96 \(\pm\) 0.22} & \textbf{89.66 \(\pm\) 0.24} & \textbf{52.22 \(\pm\) 2.30} & \textbf{81.00 \(\pm\) 0.70} & \textbf{1.431} & \textbf{14.69} \\
\makecell[l]{\textbf{TopoCryptoEEG} \\ \textbf{(ours, \(q=3\))}} & \textbf{98.39 \(\pm\) 0.41} & \textbf{6.45 \(\pm\) 0.88} & \textbf{93.06 \(\pm\) 1.19} & \textbf{73.61 \(\pm\) 7.84} & \textbf{91.77 \(\pm\) 2.01} & \textbf{2.043} & \textbf{15.11} \\
\bottomrule
\end{tabular}
\end{table*}

Under the matched single-window protocol, TopoCryptoEEG achieves AUCs of 99.50\% on HGD and 96.21\% on BCICIV-2a, with EERs of 3.15\% and 9.96\%, respectively. At FAR = 1\%, TAR reaches 91.61\% and 52.22\%. With three-window aggregation, EER further decreases to 1.21\% on HGD and 6.45\% on BCICIV-2a, while TAR@FAR=1\% increases by 6.67 and 21.39 percentage points, respectively.

TopoCryptoEEG offers substantial computational savings. When adapted to brainprints, CryptoFace suffers from modality mismatch, incurring prohibitive latency ($>$1,400 s) and memory usage (47-56 GiB). In contrast, TopoCryptoEEG achieves a per-claim encrypted-computation latency of 1.4–1.6 s ($>$18× faster than EER's~28 s) and requires only~14 GiB of memory, vastly outperforming CryptoFace.

Table \ref{tab:topofuse_baseline} shows that TopoFuse-MLP performs comparably to BrainprintNet on HGD and achieves better results than the evaluated baselines on BCICIV-2a, while retaining a compact architecture for encrypted inference.
\begin{table*}[!t]
\centering
\caption{Comparison of TopoFuse-MLP with existing state-of-the-art baseline models.}
\label{tab:topofuse_baseline}
\small
\setlength{\tabcolsep}{6pt} 
\begin{tabular}{llccccc}
\toprule
\textbf{Dataset} & \textbf{Method} & \makecell{\textbf{AUC} \\ (\%)} & \makecell{\textbf{EER} \\ (\%)} & \makecell{\textbf{BA} \\ (\%)} & \makecell{\textbf{TAR@1\%} \\ (\%)} & \makecell{\textbf{TAR@5\%} \\ (\%)} \\
\midrule
\multirow{4}{*}{HGD} 
& BrainprintNet \cite{ref38} & 99.93 \(\pm\) 0.06 & 0.96 \(\pm\) 0.51 & 98.41 \(\pm\) 0.42 & 98.90 \(\pm\) 0.83 & 99.70 \(\pm\) 0.34 \\
& ADN \cite{ref39} & 99.77 \(\pm\) 0.44 & 1.61 \(\pm\) 2.12 & 97.80 \(\pm\) 2.95 & 94.61 \(\pm\) 10.33 & 98.44 \(\pm\) 3.21 \\
& DARN \cite{ref40} & 91.96 \(\pm\) 1.46 & 16.10 \(\pm\) 1.43 & 82.07 \(\pm\) 1.52 & 38.07 \(\pm\) 2.43 & 56.70 \(\pm\) 2.42 \\
& \makecell[l]{\textbf{TopoFuse-MLP} \\ \textbf{(Ours)}} & \textbf{99.93 \(\pm\) 0.03} & \textbf{1.21 \(\pm\) 0.28} & \textbf{98.51 \(\pm\) 0.53} & \textbf{98.28 \(\pm\) 1.04} & \textbf{99.83 \(\pm\) 0.19} \\
\midrule
\multirow{4}{*}{BCICIV-2a} 
& BrainprintNet \cite{ref38} & 91.46 \(\pm\) 0.34 & 17.84 \(\pm\) 0.45 & 82.05 \(\pm\) 0.45 & 41.69 \(\pm\) 6.62 & 60.90 \(\pm\) 1.70 \\
& ADN \cite{ref39} & 66.97 \(\pm\) 10.23 & 38.48 \(\pm\) 7.06 & 64.74 \(\pm\) 5.62 & 2.90 \(\pm\) 5.86 & 20.80 \(\pm\) 9.79 \\
& DARN \cite{ref40} & 66.33 \(\pm\) 2.49 & 35.27 \(\pm\) 2.48 & 67.36 \(\pm\) 1.62 & 2.62 \(\pm\) 3.12 & 32.05 \(\pm\) 4.35 \\
& \makecell[l]{\textbf{TopoFuse-MLP} \\ \textbf{(Ours)}} & \textbf{98.39 \(\pm\) 0.41} & \textbf{6.45 \(\pm\) 0.88} & \textbf{93.06 \(\pm\) 1.19} & \textbf{73.61 \(\pm\) 7.84} & \textbf{91.77 \(\pm\) 2.01} \\
\bottomrule
\end{tabular}
\end{table*}

Table \ref{tab:decision_agreement} reports score errors and decision agreement over 40 evaluated requests. On both datasets, the maximum absolute error is below the minimum decision margin—the smallest absolute plaintext score after threshold adjustment—and no authentication decision is changed.

\begin{table}[!t]
\centering
\caption{Plaintext-ciphertext decision agreement and score errors.}
\label{tab:decision_agreement}
\begin{tabular}{lcc}
\toprule
\textbf{Metric} & \textbf{HGD} & \textbf{BCICIV-2a} \\
\midrule
Plaintext BA (\%) & 97.5 & 90.0 \\
Ciphertext BA (\%) & 97.5 & 90.0 \\
Decision agreement (\%) & 100 & 100 \\
Mean error & \(2.44 \times 10^{-6}\) & \(3.59 \times 10^{-6}\) \\
Maximum absolute error & \(9.23 \times 10^{-6}\) & \(1.02 \times 10^{-5}\) \\
Minimum decision margin & \(3.19 \times 10^{-5}\) & \(1.93 \times 10^{-5}\) \\
\bottomrule
\end{tabular}
\end{table}

Table \ref{tab:latency_resources} compares TopoCryptoEEG and FLEA. TopoCryptoEEG's GPU execution reduces encrypted-computation latency from 1.5–2.0 s (CPU) to~14 ms ($>$110× speedup). FLEA achieves a similar~13 ms latency, but only by leaving feature extraction in plaintext on the client, retaining template-recovery risks. Consequently, TopoCryptoEEG incurs only 1–2 s of CPU overhead (or matches FLEA on GPU) to provide stronger end-to-end security, effectively balancing efficiency and privacy.

\begin{table}[!t]
\centering
\caption{Online encrypted-computation latency and resource use.}
\label{tab:latency_resources}
\begin{tabular}{lccc}
\toprule
\textbf{Metric} & \textbf{Method} & \textbf{HGD} & \textbf{BCICIV-2a} \\
\midrule
\multirow{3}{*}{Mean latency (ms)} 
& CPU (Ours) & \(1578 \pm 396\) & \(2043 \pm 18\) \\
& GPU (Ours) & \(14.13 \pm 0.22\) & \(14.05 \pm 0.10\) \\
& FLEA & 13.5 & 12.9 \\
\midrule
\multirow{3}{*}{\makecell[l]{Memory / GPU \\ memory (GiB)}} 
& CPU (Ours) & 14.03 & 15.11 \\
& GPU (Ours) & 22.26 & 22.26 \\
& FLEA & 0.041 & 0.035 \\
\midrule
\multirow{3}{*}{GPU speedup} 
& CPU (Ours) & --- & --- \\
& GPU (Ours) & 111.7\(\times\) & 145.4\(\times\) \\
& FLEA & --- & --- \\
\bottomrule
\end{tabular}
\end{table}

\subsection{Ablation Study}

Table \ref{tab:ablation_plaintext} shows that each MWTE component contributes to authentication performance. Removing multi-window aggregation increases EER from 1.21\% to 3.15\% on HGD and from 6.45\% to 9.96\% on BCICIV-2a. Removing topology-aware mapping further reduces AUC from 99.50\% to 97.48\% on HGD and from 96.21\% to 89.65\% on BCICIV-2a; replacing statistical features with raw EEG causes a further substantial decline.

\begin{table*}[!t]
\centering
\caption{Progressive ablation results on the HGD and BCICIV-2a datasets.}
\label{tab:ablation_plaintext}
\small
\setlength{\tabcolsep}{6pt}
\begin{tabular}{llccccc}
\toprule
\textbf{Dataset} & \textbf{Method} & \makecell{\textbf{AUC} \\ (\%)} & \makecell{\textbf{EER} \\ (\%)} & \makecell{\textbf{BA} \\ (\%)} & \makecell{\textbf{TAR@1\%} \\ (\%)} & \makecell{\textbf{TAR@5\%} \\ (\%)} \\
\midrule
\multirow{4}{*}{HGD} 
& D: MWTE & 99.93 \(\pm\) 0.03 & 1.21 \(\pm\) 0.28 & 98.51 \(\pm\) 0.52 & 98.28 \(\pm\) 1.04 & 99.83 \(\pm\) 0.19 \\
& C: Topo\_single & 99.50 \(\pm\) 0.16 & 3.15 \(\pm\) 0.56 & 96.63 \(\pm\) 0.70 & 91.61 \(\pm\) 1.86 & 97.94 \(\pm\) 0.93 \\
& B: Stat\_no\_topo & 97.48 \(\pm\) 0.52 & 8.79 \(\pm\) 1.04 & 90.50 \(\pm\) 0.95 & 69.51 \(\pm\) 6.09 & 86.56 \(\pm\) 2.81 \\
& A: Raw EEG & 75.89 \(\pm\) 2.01 & 31.17 \(\pm\) 1.84 & 69.20 \(\pm\) 1.84 & 4.53 \(\pm\) 0.46 & 18.47 \(\pm\) 1.72 \\
\midrule
\multirow{4}{*}{BCICIV-2a} 
& D: MWTE & 98.39 \(\pm\) 0.41 & 6.45 \(\pm\) 0.88 & 93.06 \(\pm\) 1.19 & 73.61 \(\pm\) 7.84 & 91.77 \(\pm\) 2.01 \\
& C: Topo\_single & 96.21 \(\pm\) 0.11 & 9.96 \(\pm\) 0.22 & 89.66 \(\pm\) 0.24 & 52.22 \(\pm\) 2.30 & 81.00 \(\pm\) 0.70 \\
& B: Stat\_no\_topo & 89.65 \(\pm\) 1.52 & 18.99 \(\pm\) 1.47 & 81.05 \(\pm\) 1.67 & 23.45 \(\pm\) 3.52 & 50.33 \(\pm\) 5.52 \\
& A: Raw EEG & 67.07 \(\pm\) 1.82 & 38.18 \(\pm\) 1.55 & 61.69 \(\pm\) 1.87 & 3.56 \(\pm\) 1.16 & 11.77 \(\pm\) 2.83 \\
\bottomrule
\end{tabular}
\end{table*}

\begin{table*}[!t]
\centering
\caption{Ciphertext organization and logical homomorphic-operator counts per authentication claim.}
\label{tab:ablation_ciphertext}
\small
\setlength{\tabcolsep}{6pt}
\begin{tabular}{llcccccc}
\toprule
\textbf{Configuration} & \textbf{Ciphertext packing} & \textbf{Linear transformation} & \textbf{Backend} & \makecell{\textbf{Input} \\ \textbf{encryption}} & \textbf{MultiplyPlain} & \textbf{Rotate} & \textbf{Square} \\
\midrule
H0 & Independent patches & Naive & CPU & 12 & 3817 & 3219 & 27 \\
H1 & Independent patches & Standard BSGS & CPU & 12 & 3817 & 663 & 27 \\
H2 & QPCP & Naive & CPU & 1 & 261 & 243 & 3 \\
H3 & QPCP & LP-BSGS & CPU & 1 & 261 & 57 & 3 \\
H4 & QPCP & LP-BSGS & GPU & 1 & 261 & 57 & 3 \\
\bottomrule
\end{tabular}
\end{table*}

Table~\ref{tab:ablation_ciphertext} isolates the encrypted-computation optimizations. Standard BSGS reduces rotations from 3,219 to 663; QPCP reduces input encryptions from 12 to 1 and MultiplyPlain operations from 3,817 to 261; LP-BSGS then reduces rotations to 57. Standard BSGS (H1) achieves an average speedup of 6.0× over H0, whereas QPCP (H2) achieves 14.2×. Adding LP-BSGS (H3) provides a further average speedup of 4.0× over H2. GPU execution (H4) further accelerates H3 by 128.5× on average. Overall, H4 achieves average speedups of 510.6× over H2 and 7,264× over H0. (Fig.~5).

\begin{figure*}[!t]
\centering
\subfloat[]{%
\includegraphics[width=0.48\textwidth]{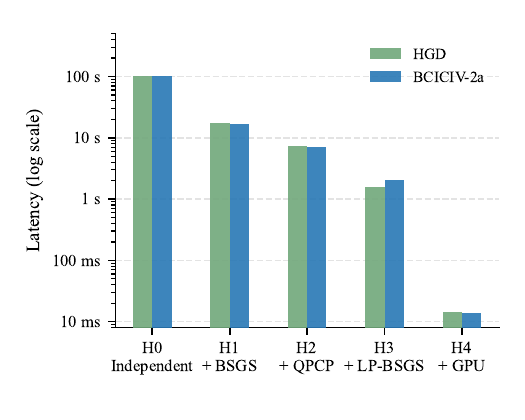}%
\label{fig_latency}}
\hfil
\subfloat[]{%
\includegraphics[width=0.48\textwidth]{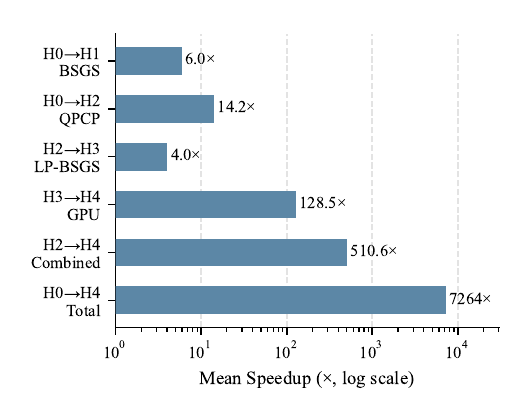}%
\label{fig_speedup}}
\caption{Encrypted-computation ablation: (a) mean online latency for H0--H4 on a logarithmic scale; (b) speedup for each configuration.}
\end{figure*}

\section{Conclusion}
We proposed and implemented TopoCryptoEEG, which, to the best of our knowledge, is the first end-to-end encrypted brainprint-authentication framework based on homomorphic encryption. After the client encrypts the data, feature extraction and all subsequent operations are performed by the server in the ciphertext domain without decryption. Because the client does not observe plaintext query embeddings, the known-feature linear template-recovery attack analyzed above cannot be directly applied. To address the high computational cost of homomorphic encryption and the complex dimensional structure of EEG, the central contribution is a co-design of Multi-Window Topology-Aware Encoding (MWTE), the Topographic Branch Model with Linear Fusion (TopoFuse-MLP), Query–Patch Coordinate Packing (QPCP), and Lane-Preserving BSGS Evaluation (LP-BSGS), balancing authentication accuracy against computational cost. Experiments on standard datasets show that TopoCryptoEEG maintains authentication accuracy while reducing per-request online encrypted-computation latency to 1–2 s on the CPU and 14 ms on the GPU, with an average speedup of more than 18\(\times\) over EER. TopoCryptoEEG offers a feasible technical path toward secure and practical brainprint-authentication systems.


\begin{IEEEbiographynophoto}{Qingyu Zeng}
is currently working toward the Ph.D. degree with Information Engineering University, under the supervision of Prof. Lei Sun. His research interests mainly include brain-computer interface security.
\end{IEEEbiographynophoto}

\begin{IEEEbiographynophoto}{Lei Sun}
received the Ph.D. degree from Information Engineering University in 2006, where he is currently a professor. He has authored or coauthored more than 60 papers in national and international journals and conference proceedings. His research interests mainly include information system security, identity authentication, and artificial intelligence security.
\end{IEEEbiographynophoto}

\begin{IEEEbiographynophoto}{Shuai Zhang}
received the M.S. degree from Information Engineering University in 2023, where he is currently working toward the Ph.D. degree under the supervision of Prof. Lei Sun. His research interests mainly include EEG decoding, biometric recognition, and unsupervised domain adaptation.
\end{IEEEbiographynophoto}

\begin{IEEEbiographynophoto}{Ting Zhang}
received the M.S. degree in computer science. She is currently a Lecturer. Her research interests include LLM security, homomorphic encryption, privacy-preserving computation, and secure biometric authentication.
\end{IEEEbiographynophoto}

\begin{IEEEbiographynophoto}{Xiuqing Mao}
received the master's degree from Information Engineering University in 2005. He is currently an associate professor, and he is currently working toward the Ph.D. degree at Information Engineering University. He has authored or coauthored more than 20 papers in national and international journals and conference proceedings. His main research interests include artificial intelligence security, brain-computer interface security, and identity authentication.
\end{IEEEbiographynophoto}

\begin{IEEEbiographynophoto}{Min Zhao}
is currently working toward the Ph.D. degree with Information Engineering University, supervised by Professor Lei Sun. Her research interests include brain-computer interface and identity authentication.
\end{IEEEbiographynophoto}

\begin{IEEEbiographynophoto}{Wenle Dong}
is currently working toward the M.S. degree with Information Engineering University, under the supervision of Prof. Lei Sun. His research interests mainly include secure biometric authentication.
\end{IEEEbiographynophoto}

\vfill


\begin{thebibliography}{1}
\bibliographystyle{IEEEtran}

\bibitem{ref1}
I. Martinovic, D. Davies, M. Frank, D. Perito, T. Ros, and D. Song, ``On the Feasibility of Side-Channel Attacks with Brain-Computer Interfaces,'' \textit{21st USENIX Security Symposium}, pp. 143--158, 2012. \url{https://www.usenix.org/conference/usenixsecurity12/technical-sessions/presentation/martinovic}

\bibitem{ref2}
M. Ienca and R. Andorno, ``Towards New Human Rights in the Age of Neuroscience and Neurotechnology,'' \textit{Life Sci. Soc. Policy}, vol. 13, Art. no. 5, 2017, doi: 10.1186/s40504-017-0050-1.

\bibitem{ref3}
Colorado General Assembly, ``HB24-1058: Protect Privacy of Biological Data,'' signed April 17, 2024, effective August 7, 2024. \url{https://leg.colorado.gov/bills/hb24-1058} (accessed October 23, 2026).

\bibitem{ref4}
M. Jiang, S. Liu, S. Han, and D. Gu, ``Biometric-Based Two-Factor Authentication Scheme Under Database Leakage,'' \textit{Theor. Comput. Sci.}, vol. 1000, Art. no. 114552, 2024, doi: 10.1016/j.tcs.2024.114552.

\bibitem{ref5}
P. L\'{o}pez-Gonz\'{a}lez, R. Arjona, R. Rom\'{a}n, and I. Baturone, ``A Facial Authentication System Using Post-Quantum-Secure Data Generated on Mobile Devices,'' in \textit{Proc. ACM MobiCom}, 2022, pp. 803--805, doi: 10.1145/3495243.3558761.

\bibitem{ref6}
Y. Zhang, H. Yi, and W. Kong, ``A Privacy-Preserving Brainprint Recognition System Based on Feature Homomorphic Encryption,'' in \textit{Proc. IJCNN}, 2024, pp. 1--10, doi: 10.1109/IJCNN60899.2024.10651340.

\bibitem{ref7}
A. Bassit, F. Hahn, Z. Rezgui, U. M. Kelly, R. Veldhuis, and A. Peter, ``Template Recovery Attack on Homomorphically Encrypted Biometric Recognition Systems with Unprotected Threshold Comparison,'' in \textit{Proc. IEEE Int. Joint Conf. Biometrics (IJCB)}, 2023, pp. 1--10, doi: 10.1109/IJCB57857.2023.10449211.

\bibitem{ref8}
E. Pagnin, C. Dimitrakakis, A. Abidin, and A. Mitrokotsa, ``On the Leakage of Information in Biometric Authentication,'' in \textit{Progress in Cryptology-INDOCRYPT 2014}, pp. 265--280, 2014, doi: 10.1007/978-3-319-13039-2\_16.

\bibitem{ref9}
G. Cha, D. Park, Y. Choi, E. Park, and J.-W. Lee, ``Real-Time Encrypted Emotion Recognition Using Homomorphic Encryption,'' \textit{Proc. ACM Interact. Mobile Wearable Ubiquitous Technol.}, vol. 9, no. 4, Art. no. 163, pp. 1--32, 2025, doi: 10.1145/3770633.

\bibitem{ref10}
R. Gilad-Bachrach, N. Dowlin, K. Laine, K. Lauter, M. Naehrig, and J. Wernsing, ``CryptoNets: Applying Neural Networks to Encrypted Data with High Throughput and Accuracy,'' in \textit{Proc. 33rd Int. Conf. Mach. Learn. (ICML)}, PMLR, vol. 48, pp. 201--210, 2016.

\bibitem{ref11}
W. Ao and V. N. Boddeti, ``CryptoFace: End-To-End Encrypted Face Recognition,'' in \textit{Proc. IEEE/CVF CVPR}, 2025, pp. 19197--19206, doi: 10.1109/CVPR52734.2025.01788.

\bibitem{ref12}
S.-X. Zhang, Y. Gong, and D. Yu, ``Encrypted Speech Recognition Using Deep Polynomial Networks,'' in \textit{Proc. ICASSP}, 2019, pp. 5691--5695, doi: 10.1109/ICASSP.2019.8683721.

\bibitem{ref13}
Y. Liu, H. Huang, F. Xiao, R. Malekian, and W. Wang, ``Classification and Recognition of Encrypted EEG Data Based on Neural Network,'' \textit{J. Inf. Security Appl.}, vol. 54, Art. no. 102567, 2020, doi: 10.1016/j.jisa.2020.102567.

\bibitem{ref14}
A. B. Popescu et al., ``Privacy Preserving Classification of EEG Data Using Machine Learning and Homomorphic Encryption,'' \textit{Appl. Sci.}, vol. 11, no. 16, Art. no. 7360, 2021, doi: 10.3390/app11167360.

\bibitem{ref15}
M. R. Abou Harb, B. Celiktas, and G. Eroglu, ``Secure and Interpretable Dyslexia Detection Using Homomorphic Encryption and SHAP-Based Explanations,'' in \textit{Proc. 2025 Medical Technologies Congress (TIPTEKNO)}, 2025, doi: 10.1109/TIPTEKNO68206.2025.11270026.

\bibitem{ref16}
H. Dong et al., ``Privacy-Preserving EEG Signal Analysis with Electrode Attention for Depression Diagnosis: Joint FHE and CNN Approach,'' in \textit{Proc. IEEE GLOBECOM}, 2023, pp. 4265--4270, doi: 10.1109/GLOBECOM54140.2023.10436783.

\bibitem{ref17}
A. Brutzkus, R. Gilad-Bachrach, and O. Elisha, ``Low Latency Privacy Preserving Inference,'' in \textit{Proc. 36th Int. Conf. Mach. Learn. (ICML)}, PMLR, vol. 97, pp. 812--821, 2019.

\bibitem{ref18}
R. Dathathri et al., ``CHET: An Optimizing Compiler for Fully-Homomorphic Neural-Network Inferencing,'' in \textit{Proc. ACM SIGPLAN Conf. Program. Lang. Design Implement.}, 2019, pp. 142--156, doi: 10.1145/3314221.3314628.

\bibitem{ref19}
R. Dathathri, B. Kostova, O. Saarikivi, W. Dai, K. Laine, and M. Musuvathi, ``EVA: An Encrypted Vector Arithmetic Language and Compiler for Efficient Homomorphic Computation,'' in \textit{Proc. ACM SIGPLAN Conf. Program. Lang. Design Implement.}, 2020, pp. 546--561, doi: 10.1145/3385412.3386023.

\bibitem{ref20}
E. Lee, J.-W. Lee, J. Lee, Y.-S. Kim, Y. Kim, J.-S. No, and W. Choi, ``Low-Complexity Deep Convolutional Neural Networks on Fully Homomorphic Encryption Using Multiplexed Parallel Convolutions,'' in \textit{Proc. 39th Int. Conf. Mach. Learn. (ICML)}, PMLR, vol. 162, pp. 12403--12422, 2022.

\bibitem{ref21}
D. Kim, J. Park, J. Kim, S. Kim, and J. H. Ahn, ``HyPHEN: A Hybrid Packing Method and Its Optimizations for Homomorphic Encryption-Based Neural Networks,'' \textit{IEEE Access}, vol. 12, pp. 3024--3038, 2024, doi: 10.1109/ACCESS.2023.3348170.

\bibitem{ref22}
S. Halevi and V. Shoup, ``Faster Homomorphic Linear Transformations in HElib,'' in \textit{Advances in Cryptology--CRYPTO 2018}, pp. 93--120, 2018, doi: 10.1007/978-3-319-96884-1\_4.

\bibitem{ref23}
M. Wang, H. El-Fiqi, J. Hu, and H. A. Abbass, ``Convolutional Neural Networks Using Dynamic Functional Connectivity for EEG-Based Person Identification in Diverse Human States,'' \textit{IEEE Trans. Inf. Forensics Security}, vol. 14, no. 12, pp. 3259--3272, 2019, doi: 10.1109/TIFS.2019.2916403.

\bibitem{ref24}
M. Wang, J. Hu, and H. A. Abbass, ``BrainPrint: EEG Biometric Identification Based on Analyzing Brain Connectivity Graphs,'' \textit{Pattern Recognit.}, vol. 105, Art. no. 107381, 2020, doi: 10.1016/j.patcog.2020.107381.

\bibitem{ref25}
Y. Du, Y. Xu, X. Wang, L. Liu, and P. Ma, ``EEG Temporal-Spatial Transformer for Person Identification,'' \textit{Sci. Rep.}, vol. 12, Art. no. 14378, 2022, doi: 10.1038/s41598-022-18502-3.

\bibitem{ref26}
S. Marcel and J. R. Millán, ``Person Authentication Using Brainwaves (EEG) and Maximum A Posteriori Model Adaptation,'' \textit{IEEE Trans. Pattern Anal. Mach. Intell.}, vol. 29, no. 4, pp. 743--752, 2007, doi: 10.1109/TPAMI.2007.1012.

\bibitem{ref27}
D. La Rocca et al., ``Human Brain Distinctiveness Based on EEG Spectral Coherence Connectivity,'' \textit{IEEE Trans. Biomed. Eng.}, vol. 61, no. 9, pp. 2406--2412, 2014, doi: 10.1109/TBME.2014.2317881.

\bibitem{ref28}
A. J. Bidgoly, H. J. Bidgoly, and Z. Arezoumand, ``Towards a Universal and Privacy Preserving EEG-Based Authentication System,'' \textit{Sci. Rep.}, vol. 12, Art. no. 2531, 2022, doi: 10.1038/s41598-022-06527-7.

\bibitem{ref29}
M. Wang, S. Wang, and J. Hu, ``Cancellable Template Design for Privacy-Preserving EEG Biometric Authentication Systems,'' \textit{IEEE Trans. Inf. Forensics Security}, vol. 17, pp. 3350--3364, 2022, doi: 10.1109/TIFS.2022.3204222.

\bibitem{ref30}
M. Wang, X. Yin, and J. Hu, ``Cancellable Deep Learning Framework for EEG Biometrics,'' \textit{IEEE Trans. Inf. Forensics Security}, vol. 19, pp. 3745--3757, 2024, doi: 10.1109/TIFS.2024.3369405.

\bibitem{ref31}
M. Wang, S. Wang, and J. Hu, ``PolyCosGraph: A Privacy-Preserving Cancelable EEG Biometric System,'' \textit{IEEE Trans. Dependable Secure Comput.}, vol. 20, no. 5, pp. 4258--4272, 2023, doi: 10.1109/TDSC.2022.3218782.

\bibitem{ref32}
M. Umair et al., ``Privacy-Preserving Dementia Classification from EEG via Hybrid-Fusion EEGNetv4 and Federated Learning,'' \textit{Front. Comput. Neurosci.}, vol. 19, Art. no. 1617883, 2025, doi: 10.3389/fncom.2025.1617883.

\bibitem{ref33}
V. N. Boddeti, ``Secure Face Matching Using Fully Homomorphic Encryption,'' in \textit{Proc. IEEE BTAS}, 2018, pp. 1--10, doi: 10.1109/BTAS.2018.8698601.

\bibitem{ref34}
J. H. Cheon, A. Kim, M. Kim, and Y. Song, ``Homomorphic Encryption for Arithmetic of Approximate Numbers,'' in \textit{Advances in Cryptology-ASIACRYPT 2017}, pp. 409--437, 2017, doi: 10.1007/978-3-319-70694-8\_15.

\bibitem{ref35}
Y. Song, Q. Zheng, B. Liu, and X. Gao, ``EEG Conformer: Convolutional Transformer for EEG Decoding and Visualization,'' \textit{IEEE Trans. Neural Syst. Rehabil. Eng.}, vol. 31, pp. 710--719, 2023, doi: 10.1109/TNSRE.2022.3230250.

\bibitem{ref36}
P. Mishra, R. Lehmkuhl, A. Srinivasan, W. Zheng, and R. A. Popa, ``Delphi: A Cryptographic Inference Service for Neural Networks,'' in \textit{Proc. USENIX Security Symp.}, 2020, pp. 2505--2522.

\bibitem{ref37}
P. Bashivan, I. Rish, M. Yeasin, and N. Codella, ``Learning Representations from EEG with Deep Recurrent-Convolutional Neural Networks,'' in \textit{Proc. ICLR}, 2016, doi: 10.48550/arXiv.1511.06448.

\bibitem{ref38}
Y. Tu, S. Li, X. Chen, and D. Wu, ``BrainprintNet: A Multiscale Cross-Band Fusion Network for EEG-Based Brainprint Recognition,'' \textit{IEEE Trans. Inf. Forensics Security}, vol. 21, pp. 2757--2768, 2026, doi: 10.1109/TIFS.2026.3672000.

\bibitem{ref39}
Z. Mu, J. Hu, and J. Min, ``EEG-Based Person Authentication Using a Fuzzy Entropy-Related Approach with Two Electrodes,'' \textit{Entropy}, vol. 18, no. 12, Art. no. 432, 2016, doi: 10.3390/e18120432.

\bibitem{ref40}
H. Liu et al., ``DARN: A Dual Attention Refinement Network for Enhancing Feature Robustness in VEP-Based EEG Biometrics,'' \textit{IEEE Trans. Inf. Forensics Security}, vol. 20, pp. 7166--7180, 2025, doi: 10.1109/TIFS.2025.3587181.

\bibitem{ref41}
Microsoft Research, ``Microsoft SEAL: Simple Encrypted Arithmetic Library,'' version 4.1.1, software. \url{https://github.com/microsoft/SEAL/tree/v4.1.1} (accessed October 23, 2026).

\bibitem{ref42}
A. \c{S}. \"{O}zcan and E. Sava\c{s}, ``HEonGPU: A GPU-Based Fully Homomorphic Encryption Library 1.0,'' \textit{Cryptology ePrint Archive, Paper 2024/1543}, 2024. [Online]. Available: \url{https://eprint.iacr.org/2024/1543}.

\bibitem{ref43}
J.-P. Bossuat et al., ``Security Guidelines for Implementing Homomorphic Encryption,'' \textit{IACR Commun. Cryptol.}, vol. 1, no. 4, 2025, doi: 10.62056/anxra69p1.

\bibitem{ref44}
R. T. Schirrmeister et al., ``Deep Learning With Convolutional Neural Networks for EEG Decoding and Visualization,'' \textit{Hum. Brain Mapp.}, vol. 38, pp. 5391--5420, 2017, doi: 10.1002/hbm.23730.

\bibitem{ref45}
C. Brunner, R. Leeb, G. R. M\"{u}ller-Putz, A. Schl\"{o}gl, and G. Pfurtscheller, ``BCI Competition 2008---Graz Data Set A,'' dataset description, 2008. [Online]. Available: \url{https://www.bbci.de/competition/iv/desc_2a.pdf}.

\end{thebibliography}
\end{document}